# Optical Wavelength Meter with Machine Learning Enhanced Precision

Gazi Mahamud Hasan[1*], Mehedi Hasan[1], Peng Liu[1], Mohammad Rad[2], Eric Bernier[2], Trevor James Hall[1]

[1]Photonic Technology Laboratory, Centre for Research in Photonics, Advanced Research Complex, University of Ottawa, 25 Templeton Street, Ottawa, K1N 6N5, ON, Canada
[2]Huawei Technologies Canada, 303 Terry Fox Drive, Kanata, K2K 3J1, ON, Canada
*ghasa102@uottawa.ca

## Abstract

Diverse applications in photonics and microwave engineering require a means of measurement of the instantaneous frequency of a signal. A photonic implementation typically applies an interferometer equipped with three or more output ports to measure the frequency dependent phase shift provided by an optical delay line. The components constituting the interferometer are prone to impairments which results in erroneous measurements. It is shown that the information to be retrieved is encoded by a three-component vector that lies on a circular cone within a three-dimensional Cartesian object space. The measured data belongs to the image of the object space under a linear map that describes the action of the interferometer. Assisted by a learning algorithm, an inverse map from the image space into the object space is constructed. The inverse map compensates for a variety of impairments while being robust to noise. Simulation results demonstrate that, to the extent the interferometer model captures all significant impairments, a precision limited only by the level of random noise is attainable. A wavelength meter architecture is fabricated on $Si_3N_4$ photonic integration platform to prove the method experimentally. Applied to the measured data, greater than an order of magnitude improvement in precision is achieved by the proposed method compared to the conventional method.

## I.  Introduction

Interferometric methods are extensively used in a diversity applications including inter alia optical communications coherent receiver front ends [1] and spectral monitoring [2]; Bragg grating sensor interrogation [3]; laser intensity and frequency fluctuation metrology [4,5]; and, fibre optic sensing of temperature [6], pressure [7], refractive index [8], and strain [9]. Common interferometric architectures involving a Mach-Zehnder interferometer [10], Michelson interferometer [5, 11] or Fabry–Perot interferometer [12] are employed to convert a phase shift provided by some sensing means to a measurable change in light intensity. Mach-Zehnder interferometers (MZI) formed by circuits of planar waveguides and couplers are particularly attractive for photonic integration.

The fundamental principle of a wavelength meter or frequency discriminator is the use of an interferometer to measure the phase difference between the original signal and a delayed replica of the signal. The delay converts a change of frequency to a change of relative phase between the original and replica signals. The conventional MZI structure has a co-sinusoidal response to the relative phase and hence the sensitivity to small frequency deviations varies over its period. In frequency discriminator applications, it is necessary to maintain a quadrature phase bias to maximise sensitivity and, in wavelength meter applications, the loss of precision at null and peak bias points is a concern.

To avoid this signal fading problem in a passive structure, Sheem [13] introduced a MZI architecture using $3 \times 3$ directional couplers to provide a three-phase output. Koo et al. [14] developed an analogue demodulation method that projects the three-phase output onto quadrature phase components from which a continuous phase is retrieved by a process involving differentiation, cross multiplication, summation and integration. A digital measurement technique was proposed by Jin et al. [15] where least squares estimation is applied to the digitized quadrature components to recover the variation in amplitude and phase. Todd et al. [3] disclosed a Bragg grating sensor interrogation system that projects the digitized three-phase output onto quadrature components from which the phase is retrieved via a digital arctangent function. Todd's original method assumes an ideal coupler. Todd et al [10] subsequently extended the method to incorporate non-ideal coupler parameters that involves a weighted linear combination of outputs to which the digital arctangent is applied. Xu et al [5] apply the extended approach to laser phase and frequency noise metrology. The characterisation of the impairments of a component in isolation is often not possible consequently, it is Todd's original method which has become the conventional method of interferometric data processing. In [16], simulation results showed that the conventional approach is superior to preceding methods for a high-power signal in a severely noisy environment.

Kleijn et al. [17] apply a complex nonlinear least square fitting procedure to calibration data provided by an InP photonic integrated circuit wavelength meter with a $3 \times 3$ MMI based MZI circuit architecture. A total of 10 parameters are extracted: 6 coupler scattering magnitudes; 3 coupler phase shifts; and 1 delay which enable the compensation of uncertain coupler transmission matrix components, interferometer delay imbalance, and photodetector responsivity. The convergence of the nonlinear least squares fitting algorithm requires good starting points for all parameters. Moreover, the parameter estimation requires knowledge of the source power used during calibration and operation modes. The sum of the output port powers of the interferometers is used for this purpose but this sum is only substantially independent of the measurand for small impairments. Applied to experimental measurements, an improvement by a factor of two to three in precision is achieved compared to the conventional method.

The observation that any two outputs of a delay interferometer trace out a Lissajous figure as the frequency is scanned has motivated several researchers to apply a curve fitting method developed by Fitzgibbon et al. [18] which is a specific case of Bookstein's conic-section fitting method [19]. An ellipse [11] or squircle [20] is fitted to scattered data to extract the parameters required to retrieve the phase information. A 4 port 90º hybrid coupler, long used in optical communications as the front-end of a coherent optical receiver to provide in-phase and quadrature-phase demodulation [1], has been applied more recently to the measurement of laser frequency fluctuations [4]. Recently, Chen et al. [21] demonstrated a parallel arrangement of 90º hybrid-based wavelength meters with waveguides delay lines engineered to relax temperature sensitivity.

In this paper, a data processing method is proposed that corrects the same comprehensive set of impairments as Kleijn's method while eliminating its deficiencies. The algorithm is simple and robust; substantially requiring linear algebra operations only. Only the time delay parameter requires bracketing over a broad interval; no starting points are required for all parameters. The calibration and phase retrieval process are invariant to source power. The retrieval process is naturally invariant to source optical power fluctuations during data processing. A source optical power monitor may be used advantageously to render the calibration process similarly invariant to source power fluctuations.

A complete theoretical formulation of the problem is presented in Section II for any number $n > 3$ of output ports. In Section II.A the treatment of the interferometer is restricted to perfect components. The transmission the output coupler is then described by a Fourier matrix. A real three-component object vector composed of an in-phase, quadrature phase, and input power emerges as a representation of the autocorrelation of the input sequence provided by the two arms of the interferometer. It is found that the vector belongs to a double napped circular cone within an object space $\mathbb{R}^3$. Each point on the cone is sent by an orthogonal map to a vector of interferometer egress port photo-receiver outputs in an image space $\mathbb{R}^n$ $n \geq 3$. The transpose of this orthogonal map is the basis of Todd's method of retrieving the phase from the measured output port powers. In Section II.B the restriction to perfect components is lifted. An analysis is conducted that considers intensity fluctuations of the source; the impairments of the waveguide delay line and the couplers that compose the interferometer; and the noise and sensitivity errors of the photo-receiver array. It is found that the circular cone is retained as the fundamental object on which the data to be retrieved is known to be located. The component impairments break the orthogonal symmetry but the map from the cone to the image space remains linear and, if the impairments are not too severe, it is close to orthogonal. In section II.C the information retrieval problem is formulated for a known interferometer delay as the construction of a $3 \times n$ matrix representing the linear map that minimises the sum of the squared prediction error over a training data set which is solved analytically using linear algebra only. An uncertain delay introduces nonlinearity but a modest number of iterations of the robust golden search algorithm starting from a loose bracket suffice to retrieve the delay parameter.

The theoretical predictions are verified by the results of simulations presented in Section III where it is demonstrated that a precision limited only by the level of random noise is attainable to the extent that the model of the interferometer captures all significant impairments. Section IV provides experimental proof of the proposed method. A $Si_3N_4$ photonic integrated circuit wavelength meter with a $3 \times 3$ MMI based MZI circuit architecture fabricated using the LioniX International foundry is used to provide raw data which is processed by the proposed method. Greater than an order of magnitude improvement in precision is achieved by the proposed method compared to the conventional method. Section V concludes the paper by providing a summary of its principal findings.

## II.  Theory

### A.  Perfect components

Figure 1 illustrates a dual MZI approach to the elimination of the signal fading problem suffered by a single MZI architecture [22]. The signal is split between two parallel MZIs that are notionally identical with the exception that one MZI is biased in quadrature relative to the other. Ideally each MZI is lossless and consequently the intensity of their two output ports are complementary.

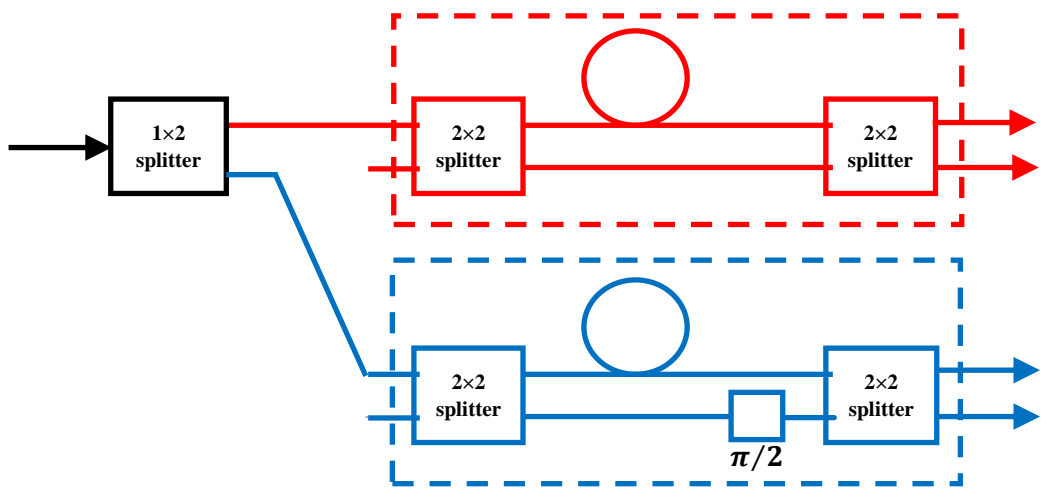

*Figure 1 Schematic of a two-stage interferometer architecture consisting of two parallel 2×2 Mach-Zehnder interferometers (MZI). The two MZI including the delay lines represented by the circles are notionally identical except for the quadrature bias of the lower (blue) MZI provided by the $\pi/2$ phase shift.*

The difference in intensity between the two output ports of each MZI provides a signed in-phase component and signed quadrature-phase component of the phasor that describes the interference term. It is then straightforward to recover the phase with a frequency invariant sensitivity.

An improvement of the architecture in which a single delay line is shared between two MZIs is illustrated in Figure 2. The sharing guarantees that the two delay lines in the equivalent architecture of Figure 1 are identical. The network of four $2 \times 2$ couplers and a $\pi/2$ phase shift is recognised as an implementation of a $4 \times 4$ discrete Fourier transform (DFT) which may be alternatively implemented using a single $4 \times 4$ coupler. For example, multimode interference (MMI) couplers with a uniform split ratio have transmission matrices that are phase-permutation equivalent to a Fourier matrix.

This rearrangement provides the motivation to employ a general interferometer architecture consisting of a $1 \times m$ uniform splitter and a $n \times n$ Fourier coupler interconnected by $m$ arms with imbalanced phase to retrieve the phasor from a datum with high precision. The transmission matrix $\boldsymbol{F} \in \mathbb{C}^{n \times n}$ of the output coupler maps a column vector $\boldsymbol{b} \in \mathbb{C}^n$ composed of the complex field amplitudes at its ingress ports to a column vector $\boldsymbol{c} \in \mathbb{C}^n$ composed of the complex field amplitude at its egress ports:

$$\boldsymbol{c} = \boldsymbol{F}\boldsymbol{b}$$

*Equation 1*

Each datum is a vector with elements equal to the modulus squared of the amplitude at each egress port, which can be identified with the diagonal of the outer product:

$$\boldsymbol{C} = \boldsymbol{c}\boldsymbol{c}^\dagger$$

*Equation 2*

which is related to the outer product $\boldsymbol{B} = \boldsymbol{b}\boldsymbol{b}^\dagger$ by:

$$\boldsymbol{C} = \boldsymbol{F}\boldsymbol{B}\boldsymbol{F}^\dagger$$

*Equation 3*

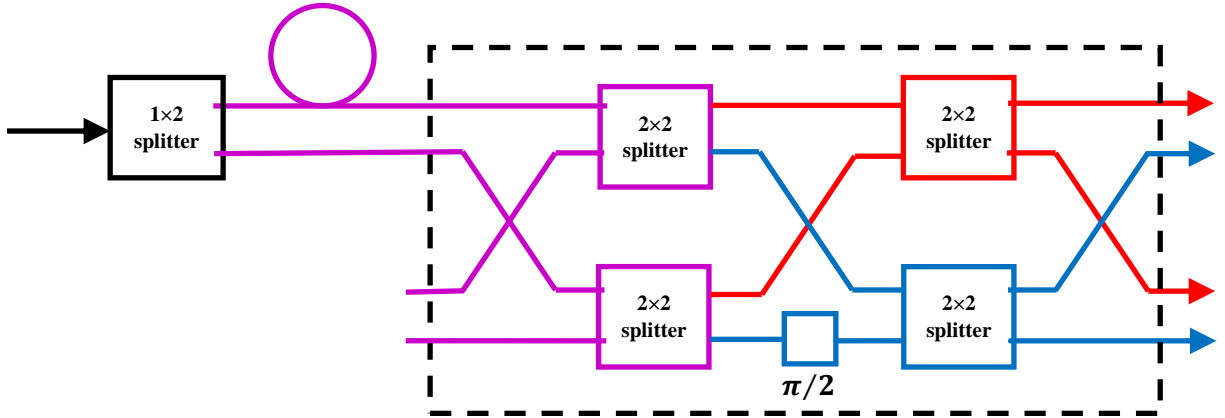

*Figure 2 A rearrangement of the interferometer architecture of Figure 1. The notionally identical arms of the two MZI excluding the phase shift have been brought forward and are now shared. The dashed subsystem block is recognised as the decomposition of a 4×4 DFT into a network of four 2×2 DFT blocks and a phase shift element.*

The measured output port power vector $\boldsymbol{p}$ is then given by:

$$\boldsymbol{p} = \sum_{j=0,n-1} \text{tr}\left(\boldsymbol{F}\boldsymbol{B}\boldsymbol{F}^\dagger \boldsymbol{e}_j \boldsymbol{e}_j^\dagger\right) \boldsymbol{e}_j = \sum_{j=0,n-1} \text{tr}\left(\boldsymbol{B}\boldsymbol{F}^\dagger \boldsymbol{e}_j \boldsymbol{e}_j^\dagger \boldsymbol{F}\right) \boldsymbol{e}_j$$

*Equation 4*

where the basis vector $\boldsymbol{e}_j$ has a unit element in position $j$ and zeros elsewhere. In the special case of a transmission matrix $\boldsymbol{F}$ that is a Fourier matrix:

$$F = \frac{1}{\sqrt{n}} \begin{bmatrix} w^0 & w^0 & \cdots & w^0 \\ w^0 & w^1 & \cdots & w^{n-1} \\ \vdots & \vdots & \ddots & \vdots \\ w^0 & w^{n-1} & \cdots & w^{(n-1)^2} \end{bmatrix} \quad ; \quad w = \exp(-i\,2\pi/n) \quad ; \quad F^\dagger F = I$$

*Equation 5*

where $I$ is the identity matrix. The outer product $B$ has the representation:

$$B = \sum_{j,k=0,1,\cdots(n-1)} b_j b_k^* \, e_j e_k^\dagger$$

*Equation 6*

Substituting Equation 6 into Equation 4 noting:

$$e_k^\dagger F^\dagger e_p e_p^\dagger F e_j = \left(e_p^\dagger F e_k\right)^\dagger \left(e_p^\dagger F e_j\right) = \frac{1}{n} w^{p(j-k)}$$

*Equation 7*

and collecting terms with $j - k = q \bmod n$, yields:

$$p = F\rho \quad ; \quad \rho_q = \frac{1}{\sqrt{n}} \sum_{j-k = q \bmod n} b_j b_k^*$$

*Equation 8*

Equation 8 is a restatement in matrix / column vector form of the familiar result that the modulus squared of the discrete Fourier transform of a sequence is equal to the discrete Fourier transform of the circular autocorrelation of the sequence. The elements of the autocorrelation vector $\rho$ are the result of summing over the trailing diagonals of $B$. A vector $b$ of length $m$ generates $2m - 1$ non-zero trailing diagonals. The cyclic nature of the summation Equation 8 does not come into play if zero padding leads to $n \geq 2m - 1$. The vector $\rho$ may then be expressed by a total of $2m - 1$ real-valued components, which is the largest number of knowable unknowns that may be recovered from the measurement. For $m = 2$, and unit input power, the vector $\rho$ takes the form:

$$\rho = \frac{1}{\sqrt{2}} \frac{1}{\sqrt{n}} \left(\cos(\theta)\,\rho_1 + \sin(\theta)\,\rho_2 + \sqrt{2}\rho_3\right)$$

*Equation 9*

where $\theta$ is the phase imbalance of the two arms, $\{\rho_1, \rho_2, \rho_3\}$ are orthonormal vectors:

$$\rho_1 = \frac{1}{\sqrt{2}} \begin{bmatrix} 0 \\ 1 \\ 0 \\ 1 \end{bmatrix} \quad \rho_2 = \frac{1}{\sqrt{2}} \begin{bmatrix} 0 \\ i \\ 0 \\ -i \end{bmatrix} \quad \rho_3 = \begin{bmatrix} 1 \\ 0 \\ 0 \\ 0 \end{bmatrix} \quad ; \quad (\rho_j, \rho_k) = \delta_{jk}$$

*Equation 10*

and $\mathbf{0}$ is the null vector representing the zero padding. The vector $\rho$ contains all the information which needs to be retrieved: the in-phase term $\cos(\theta)$, the quadrature-phase term $\sin(\theta)$, and the input power 1 appear as weights of its three orthonormal components. The phase may be extracted by introducing the scalar co-ordinates:

$$x = (\rho, \rho_1) \quad ; \quad y = (\rho, \rho_2) \quad ; \quad z = (\rho, \rho_3)$$
$$\Rightarrow$$
$$(x, y, z) = \frac{1}{\sqrt{2}} \frac{1}{\sqrt{n}} \left(\cos(\theta), \sin(\theta), \sqrt{2}\right)$$

*Equation 11*

and evaluating the arctangent:

$$\theta = \tan^{-1}\left(\frac{y}{x}\right)$$

interpreted in the four-quadrant sense. The coordinates $(x, y, z)$ satisfy the equation:

*Equation 12*

$$x^2 + y^2 - \frac{1}{2}z^2 = 0$$

*Equation 13*

which implicitly defines a double-napped circular cone in $\mathbb{R}^3$.

The Fourier matrix of Equation 5 is unitary and consequently preserves the inner product so that:

$$\boldsymbol{p} = \frac{1}{\sqrt{2}}\frac{1}{\sqrt{n}}\left(\cos(\theta)\,\boldsymbol{p}_1 + \sin(\theta)\,\boldsymbol{p}_2 + \sqrt{2}\,\boldsymbol{p}_3\right) \quad ; \quad (\boldsymbol{p}_j, \boldsymbol{p}_k) = \delta_{jk}$$

*Equation 14*

where $\{\boldsymbol{p}_1, \boldsymbol{p}_2, \boldsymbol{p}_3\}$ are the transformed basis:

$$\boldsymbol{p}_1 = \frac{\sqrt{2}}{\sqrt{n}}\begin{bmatrix}\cos(\varphi_0)\\\cos(\varphi_1)\\\vdots\\\cos(\varphi_{n-1})\end{bmatrix} \quad ; \quad \boldsymbol{p}_2 = \frac{\sqrt{2}}{\sqrt{n}}\begin{bmatrix}\sin(\varphi_0)\\\sin(\varphi_1)\\\vdots\\\sin(\varphi_{n-1})\end{bmatrix} \quad ; \quad \boldsymbol{p}_3 = \frac{1}{\sqrt{n}}\begin{bmatrix}1\\1\\\vdots\\1\end{bmatrix} \quad ; \quad \varphi_p = p\frac{2\pi}{n}$$

*Equation 15*

Together, Equation 9 and Equation 14 define a real orthogonal map $\boldsymbol{O}: \mathbb{R}^3 \to \mathbb{R}^n$ $n \geq 3$ such that:

$$\boldsymbol{p} = \frac{1}{\sqrt{2}}\frac{1}{\sqrt{n}}\boldsymbol{O}\boldsymbol{x} \quad ; \quad \boldsymbol{O} = [\boldsymbol{p}_1 \quad \boldsymbol{p}_2 \quad \boldsymbol{p}_3] \in \mathbb{R}^{n\times 3} \quad ; \quad \boldsymbol{x} = \begin{bmatrix}x\\y\\z\end{bmatrix} \in \mathbb{R}^3$$

*Equation 16*

The transpose of the orthogonal map $\boldsymbol{O}^T$ may be used to project the measured data onto the three-dimensional space $\mathbb{R}^3$ containing the circular cone. The image $\boldsymbol{p}$ of the object $\boldsymbol{x}$ under the orthogonal map also lives on a cone. The conic shape is a consequence of linearity and the absence of deformation is a consequence of orthogonality. The invariance of the cone to rotation about its axis corresponds to a translation of the phase. The mirror symmetry in any plane containing the axis or in the plane at the origin perpendicular to the axis results in a reversal of the direction clockwise or anticlockwise of increasing phase. A specific choice of coordinate system and a calibration measurement is necessary to fix the phase origin and direction.

B.   Imperfect components

To investigate the effect of impairment, the optical system can be considered equivalent to a parallel arrangement of $n$ copies of a single input and output port MZI terminated by photo-receivers and driven by a perfect $1 \times n$ splitter. Between copies, the impairments of the input splitter and interferometer arms will be in common, but the impairments of the output combiner and photo-receivers will differ. Consequently, the measurement at each egress port of the output coupler is of the form:

$$p = |a_1 \exp(i\theta) + a_2|^2 = |a_1|^2 + |a_2|^2 + 2|a_1||a_2|\cos(\theta - \phi)$$
$$\Rightarrow$$
$$p = \begin{bmatrix}2|a_1||a_2|\cos(\phi) & 2|a_1||a_2|\sin(\phi) & \frac{1}{\sqrt{2}}(|a_1|^2 + |a_2|^2)\end{bmatrix}\begin{bmatrix}\cos(\theta)\\\sin(\theta)\\\sqrt{2}\end{bmatrix}$$

*Equation 17*

where $a_1$ and $a_2$ are the complex transmission of each path excluding the phase contributed by the delay line and scaled by a real constant to account for photo-receiver sensitivity. The impairments effect the power bias $|a_1|^2 + |a_2|^2$, phase origin $\phi = \arg(a_1) - \arg(a_2)$ and amplitude $2|a_1||a_2|$ only of the underlying fringe patterns recorded by each channel while preserving the linear independence of the set of equations generated by Equation 17. The system is described by a general linear map rather than an orthogonal map of the object space to the image space, which accounts for impairments contributed by the input coupler, interferometer arms, output coupler and photo-receivers. The perfect circular cone is retained as the object space and the phase it defines relative to some origin remains meaningful as the phase contributed by the delay line. Although the orthogonal symmetry is broken, it can be expected that the linear map remains close to an orthogonal map.

## C. Learning algorithm

A linear system that maps an input $x \in \mathbb{R}^m$ to an output $y \in \mathbb{R}^m$ may be described by a matrix $A \in \mathbb{R}^{n \times m}$:

$$y = Ax$$

*Equation 18*

Suppose a sequence of measurements is made of pairs of inputs and outputs associated by the system which are assembled into a collection of data $\mathcal{D}$ known as the training set:

$$\mathcal{D} = \{(x_k, y_k) | k = 1, N\}$$

*Equation 19*

The task is to reconstruct $A$ from $\mathcal{D}$. In practice the training set $\mathcal{D}$ is corrupted by measurement errors and noise so the problem is reformulated as finding an $A$ that minimises an error function defined by:

$$F(A) = \frac{1}{N} \sum_{k=1,N} (y_k - Ax_k, y_k - Ax_k)$$

*Equation 20*

where $(\cdot,\cdot)$ denotes the Frobenius inner product. The Gâteau derivative of $F$ evaluated on the tangent vector $h$ is given by:

$$D_A F(h) = -2(h, R_{yx} - A R_{xx})$$

*Equation 21*

where:

$$R_{yx} = \frac{1}{N} \sum_{k=1,N} y_k x_k^\dagger \quad ; \quad R_{xx} = \frac{1}{N} \sum_{k=1,N} x_k x_k^\dagger$$

*Equation 22*

Consequently, the error function is minimised by the choice:

$$A = R_{yx} R_{xx}^{-1}$$

*Equation 23*

In general, $A$ is not invertible if $n > 3$. However, the Moore-Penrose inverse $A^+$ provides the minimum norm least squares solution of Equation 18. In practice, the system is over determined which leads to the explicit expression:

$$A^+ = (A^\dagger A)^{-1} A^\dagger$$

*Equation 24*

An individual measurement $y$ can be mapped to the object space by evaluating:

$$x = (A^+ y, e_1) \quad ; \quad y = (A^+ y, e_2) \quad ; \quad z = (A^+ y, e_3)$$

*Equation 25*

and its phase retrieved using $\theta = \tan^{-1}(x/y)$, where the arc tangent function is interpreted in the four-quadrant sense. However, in experiments, there is no direct object space measurement, only the frequency of the input is measured and paired with the image space measurement. The object space data is parameterised by the phase $\theta$ that must be inferred from the measured frequency $\omega$ using, if dispersion is neglected, the affine relationship:

$$\theta = (\omega - \omega_o)\tau + \theta_0$$

*Equation 26*

where $\omega_o$ is some nominal reference frequency, $\tau$ is the interferometer delay imbalance, and $\theta_0$ is the phase at the nominal reference frequency.

The phase bias in Equation 26 is sensitive to fabrication process variations and hence uncertain. Since an optical path length error of one wavelength is enough to change $\theta_0$ by $2\pi$, *a priori* $\theta_0$ can take any value in the range $(-\pi, \pi]$ albeit the precise value should be stable given a sufficiently stable ambient environment. The phase bias acts as a rotation about its axis of the circular cone on which the object samples live. The group of rotations is a subgroup of the general linear group to which ***A*** belongs. Consequently, the phase bias in Equation 26 may be dropped and its action as a rotation absorbed into ***A***.

The delay is robust to fluctuations of the ambient environment. It may be determined by design through an accurate knowledge of the physical path length imbalance and the group index of the waveguide and refined by a measurement of the free-spectral range of the interferometer. The later may be done by applying a golden section search for the delay $\tau$ that minimises the residual error given by Equation 20 after substitution of Equation 23.

## III. Simulation

A schematic of the conventional wavelength interrogation system considered for validation of the proposed method is shown in Figure 3(a). No adjustment of the proposed data processing method is necessary because of the phase permutation equivalence to a Fourier matrix of the ideal MMI transmission matrices. The unbalanced MZI architecture consists of a $2 \times 2$ MMI input coupler and $3 \times 3$ MMI output coupler with an ideal path length difference between its two arms corresponding to a free spectral range (FSR) of 50 GHz. Figure 3(b) shows the optical spectra at the three outputs of the $3 \times 3$ MMI. The Virtual Photonics Inc. (VPI) software package has been used for the simulation of the ideal system.

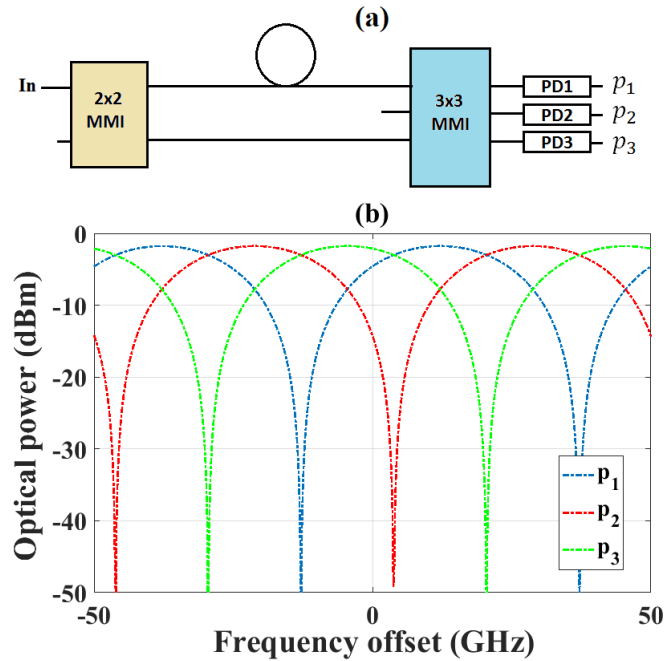

*Figure 3(a) Schematic of a conventional wavelength meter system, (b) Ideal optical spectra of the outputs of the output coupler. MMI: Multimode interference coupler, PD: Photodetector*

For an ideal system, the outputs of the identical photo-receivers can be expressed as:

$$p_1 = \frac{K}{3}\left[1 + \cos\left(\theta + \frac{5\pi}{6}\right)\right] \quad ; \quad p_2 = \frac{K}{3}\left[1 + \cos\left(\theta - \frac{3\pi}{6}\right)\right] \quad ; \quad p_3 = \frac{K}{3}\left[1 + \cos\left(\theta + \frac{\pi}{6}\right)\right]$$

$$\Rightarrow$$

$$\boldsymbol{p} = \begin{bmatrix} p_1 \\ p_2 \\ p_3 \end{bmatrix} = \frac{1}{\sqrt{2}}\frac{K}{\sqrt{3}}\boldsymbol{O}\begin{bmatrix} \cos(\theta) \\ \sin(\theta) \\ \sqrt{2} \end{bmatrix}$$

*Equation 27*

where $K$ is proportional to the input optical power and the responsivity of the photo-receivers, and:

$$\boldsymbol{O} = \frac{1}{\sqrt{3}}\begin{bmatrix} -\frac{\sqrt{3}}{\sqrt{2}} & -\frac{1}{\sqrt{2}} & 1 \\ 0 & \sqrt{2} & 1 \\ \frac{\sqrt{3}}{\sqrt{2}} & -\frac{1}{\sqrt{2}} & 1 \end{bmatrix} \quad ; \quad \boldsymbol{O}^T\boldsymbol{O} = \boldsymbol{I}$$

*Equation 28*

Equation 27 and Equation 28 and may be derived from Equation 14, Equation 15 and Equation 16 with appropriate allowance for phase permutation equivalence of MMI and Fourier couplers. Impairments due to imperfect couplers, photo-receivers and interferometer arms break the orthogonal symmetry, but the concept of the circular cone as the fundamental object remains useful since all these impairments are encompassed by the linear mapping $\boldsymbol{A}$. Fluctuations and noise will also be added by source power fluctuations, photo-receiver and quantization noise. These errors are accommodated by the least squares fitting of the system map to the training set and the Moore-Penrose inverse used for data processing. The deviations from the ideal case will be small and the port data clearly recognisable as a poly-phase fringe pattern. The phase retrieved for a given linear map is robust to source power fluctuations as the linearity in input power of the system ensures that the object samples lie on the cone irrespective of source power.

MATLAB code was developed to evaluate the performance of the learning algorithm in processing data generated by a simulated wavelength meter subject to a variety of random impairments. Object space samples are generated with perturbed interferometer delay imbalance and arbitrary phase bias. Impairments are added to the coupler transmission matrices and to the responsivities to emulate fabrication process variations and component tolerances. Gaussian noise is added to the measured data to emulate thermal and quantisation noise processes. A training set of data is generated for calibration which is used to estimate and refine the delay imbalance and thus obtain $\boldsymbol{A}$. Table 1 lists the simulation parameters and the values used.

*Table 1 Simulation parameters*

| Couplers | Gaussian distributed real & imaginary parts of transmission matrix components | symmetry preserving perturbation $\sigma = 10\%$ |
| --- | --- | --- |
| | | symmetry breaking perturbation $\sigma = 1\%$ |
| Delay line | Gaussian distributed delay | $\sigma = 5\%$ |
| Photoreceiver | Gaussian distributed responsivity | $\sigma = 10\%$ |
| Noise | Additive Gaussian noise | $\sigma = 5 \times 10^{-4}\ mW$ (source power 1mW) |

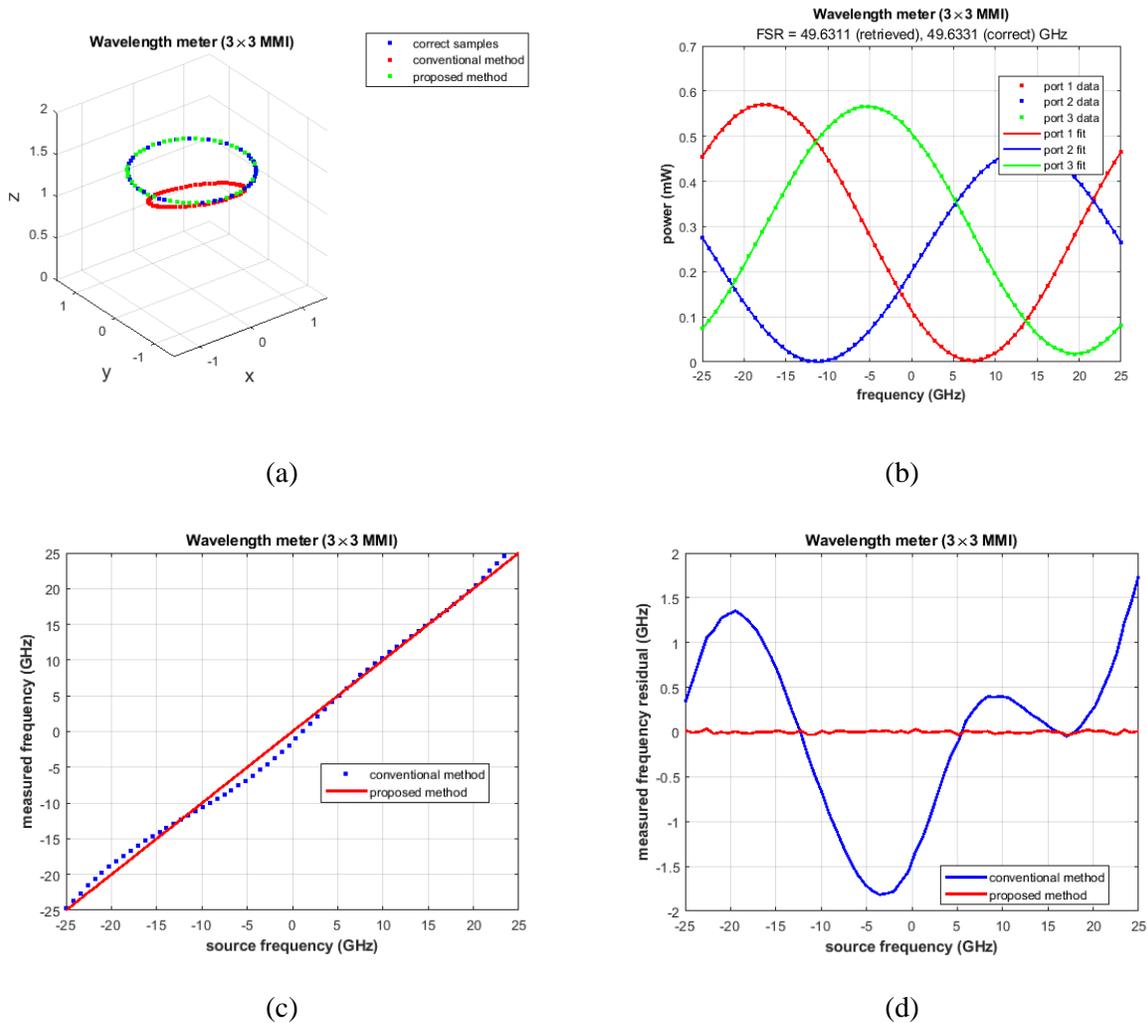

*Figure 4 (a) Correct object samples, object samples retrieved by the conventional method and object samples retrieved using the proposed method; (b) the output port fringe pattern samples (marker) accompanied by the fitted fringe pattern (solid) provided by the proposed method; (c) a comparison between the frequency measured using the conventional and proposed methods; (d) a comparison between the residual between measured and source frequency using the conventional and proposed methods. The wavelength meter simulated has a MZI architecture based on a $3 \times 3$ MMI output coupler with all components impaired.*

The simulation trials confirmed that:

1. The training and phase retrieval algorithms are invariant to static source power and phase bias. The calibration and test source powers may differ in value. The retrieved phase is naturally invariant to fluctuations from sample to sample of the test source power. The calibration process with a source power monitor is also invariant to fluctuations from sample to sample of the calibration source power.
2. The code functions with a training set containing as few as 4 frequency samples for both the $3 \times 3$ MMI and the $4 \times 4$ MMI. For the $3 \times 3$ MMI, the 4 frequency samples provide 12 knowns to retrieve 10 unknown parameters. For the $4 \times 4$ MMI, the 4 frequency samples provide 16 knowns to retrieve 13 unknown parameters.
3. If there is no random additive noise, the proposed method fully corrects all simulated impairments to machine precision provided the golden section search accuracy parameter is small enough. The loss of precision with increasing additive noise is graceful. The largest contributor to the loss of precision is additive noise in the phase retrieval process. The loss of precision of the calibration process due to additive noise is reduced by the averaging over the training set. It is expected that the additive noise in most applications will be small (SNR > 30 dB) given the modest photo-receiver bandwidth requirement.

Figure 4 shows the associated simulation results for an impaired version of the wavelength interrogation system shown in Figure 3(a). Figure 4(a) shows that the projection by the Moore-Penrose inverse $A^+$ of the simulated measured data (Figure 4(b)) has an excellent match to the original object data. Likewise the mapping of the original object space by the linear map $A$ estimated from the training data provides an excellent fit to the simulated output port fringe patterns shown in Figure 4(b). To judge the efficacy of the proposed algorithm, the conventional method due to Todd et al [3], where impaired image space data is processed by the orthogonal mapping of the perfect interferometer, has also been applied. It can be observed from Figure 4(a) that the conventional method results in poor object sample estimation which is reflected in the corresponding retrieved frequency plot shown in Figure 4(c) & (d). A substantial improvement in the accuracy of the retrieved frequency is achieved by proposed algorithm, as shown in Figure 4(c) & (d).

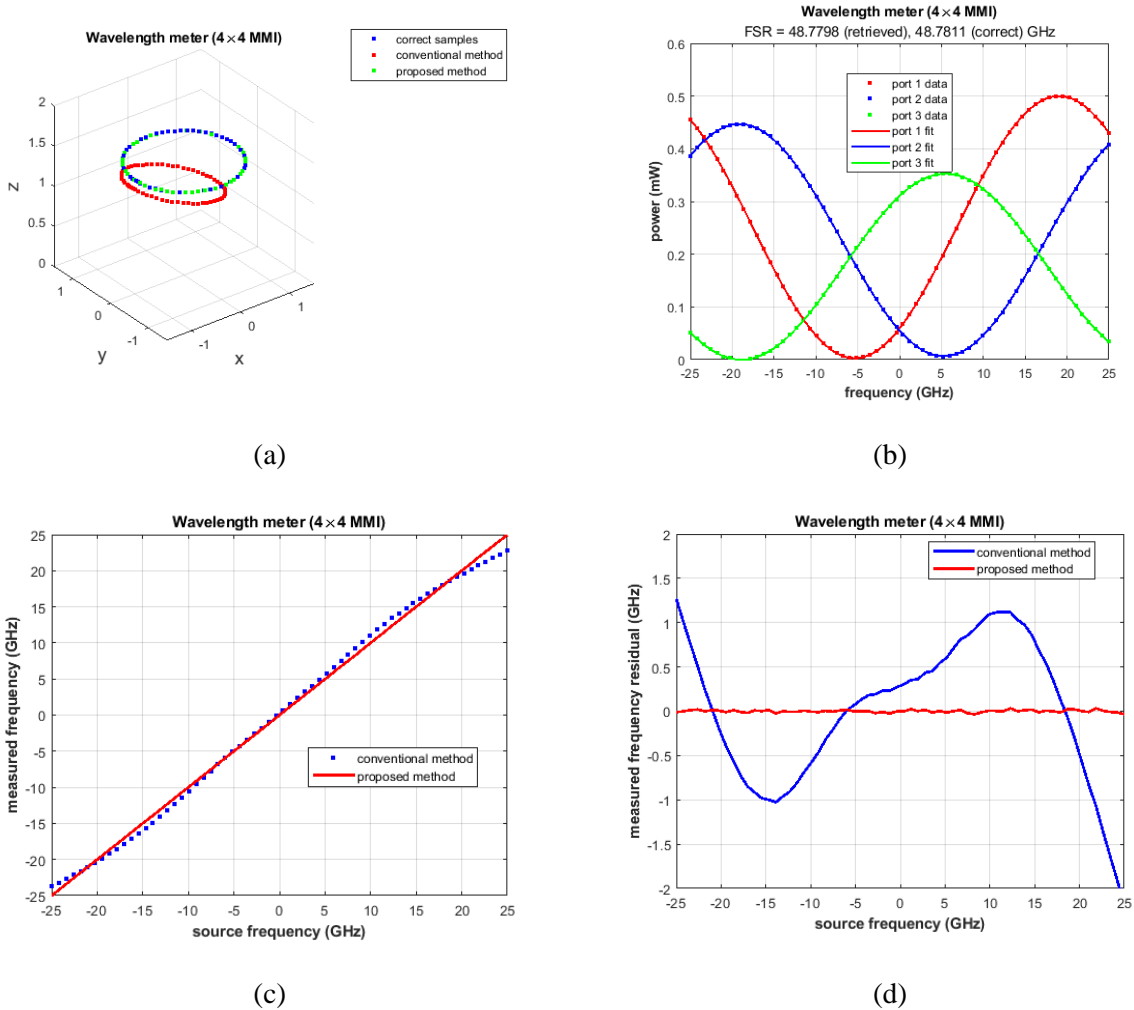

*Figure 5 (a) Correct object samples, object samples retrieved by the conventional method and object samples retrieved using the proposed method; (b) the output port fringe pattern samples (marker) accompanied by the fitted fringe pattern (solid) provided by the proposed method; (c) a comparison between the frequency measured using the conventional and proposed methods; (d) a comparison between the residual between measured and source frequency using the conventional and proposed methods. The wavelength meter simulated has a MZI architecture based on a 4 × 4 MMI output coupler with all components impaired.*

To check the generality of the proposed algorithm, the 3×3 MMI coupler is replaced by a 4×4 MMI coupler with two ingress ports left unused. The resulting orthogonal map is:

$$O = \frac{1}{\sqrt{4}} \begin{bmatrix} -1 & 1 & 1 \\ -1 & -1 & 1 \\ 1 & 1 & 1 \\ 1 & -1 & 1 \end{bmatrix} \quad ; \quad O^T O = I$$

*Equation 29*

The conventional and proposed algorithms have been applied to the same set of impaired four-dimensional image space data. The results shown in Figure 5 validate the superior accuracy of the proposed algorithm.

## IV. Experiment

To evaluate the efficacy of the proposed data processing method, experimental data is provided by a photonic integrated circuit wavelength meter with a $3 \times 3$ MMI based MZI circuit architecture fabricated on the CMOS compatible $Si_3N_4$ photonic integration platform provided by LioniX International. Their TriPlex technology offers a variety of planar waveguide structures based on alternating silicon nitride and silicon dioxide films [23]. Among them, only the asymmetric double strip (ADS) waveguide is offered by their multi-project wafer (MPW) service. The development of an on-chip wavelength meter on $Si_3N_4$ was motivated by research on a compact high-resolution wideband spectrometer [24]. To meet the specifications such as low loss, low dispersion, <1 GHz resolution, whole C band operation and compact size for the spectrometer, ADS technology on $Si_3N_4$ was chosen as the most suitable option.

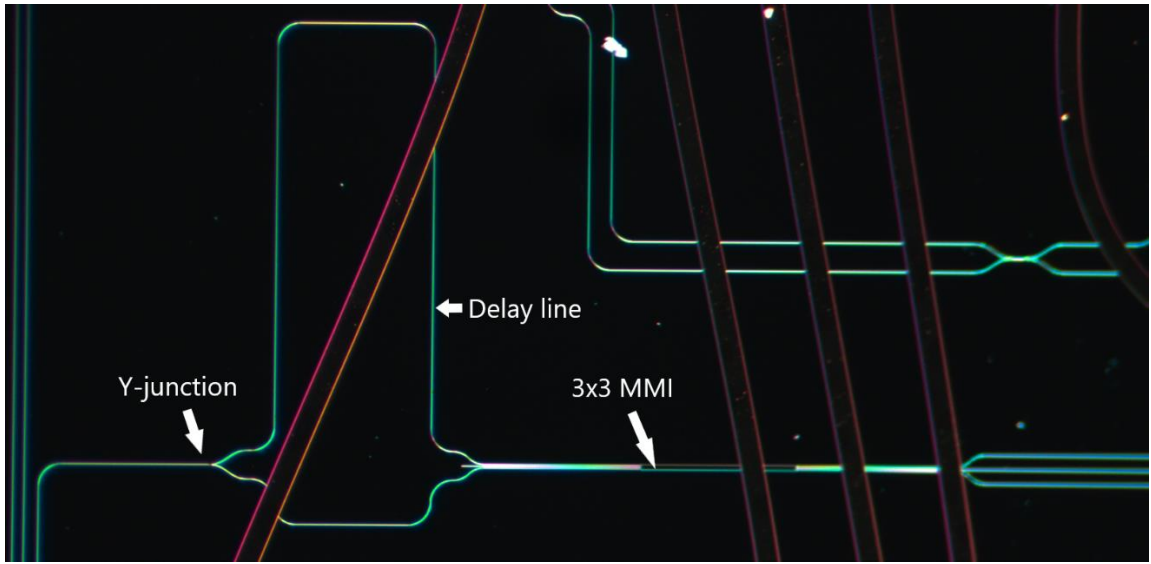

*Figure 6 Micrograph of the fabricated on-chip wavelength meter*

Figure 6 shows the micrograph of the fabricated circuit. The MZI architecture consists of a Y-junction as the input coupler and a $3 \times 3$ MMI as the output coupler with a path length difference between its arms of 3393 $\mu m$. The associated FSR for the ADS waveguide is ~49.69 GHz at the reference wavelength 1.55 μm (193.4 THz). Each input and output waveguide is terminated via spot size converter (SSC) and attached optical fibre which are not shown in Figure 6. The ADS waveguide is optimized for TE mode propagation so polarization maintaining fibers with principal axes aligned with the chip are employed. A tunable laser (Agilent 81680A) capable of tuning over the whole C-band with 3 pm wavelength step is used as the optical input. The input power is fixed at 0 dBm. The wavelength response of the circuit is measured for a desired wavelength span around 1.55 $\mu m$. The output is detected by optical power sensor (Agilent 81632A) and recorded by a lightwave measurement system (Agilent 8164A). The optical spectral data is collected and processed off-line by the proposed method, as described in Section II.C. The experiment has been conducted in a centrally temperature-controlled laboratory environment.

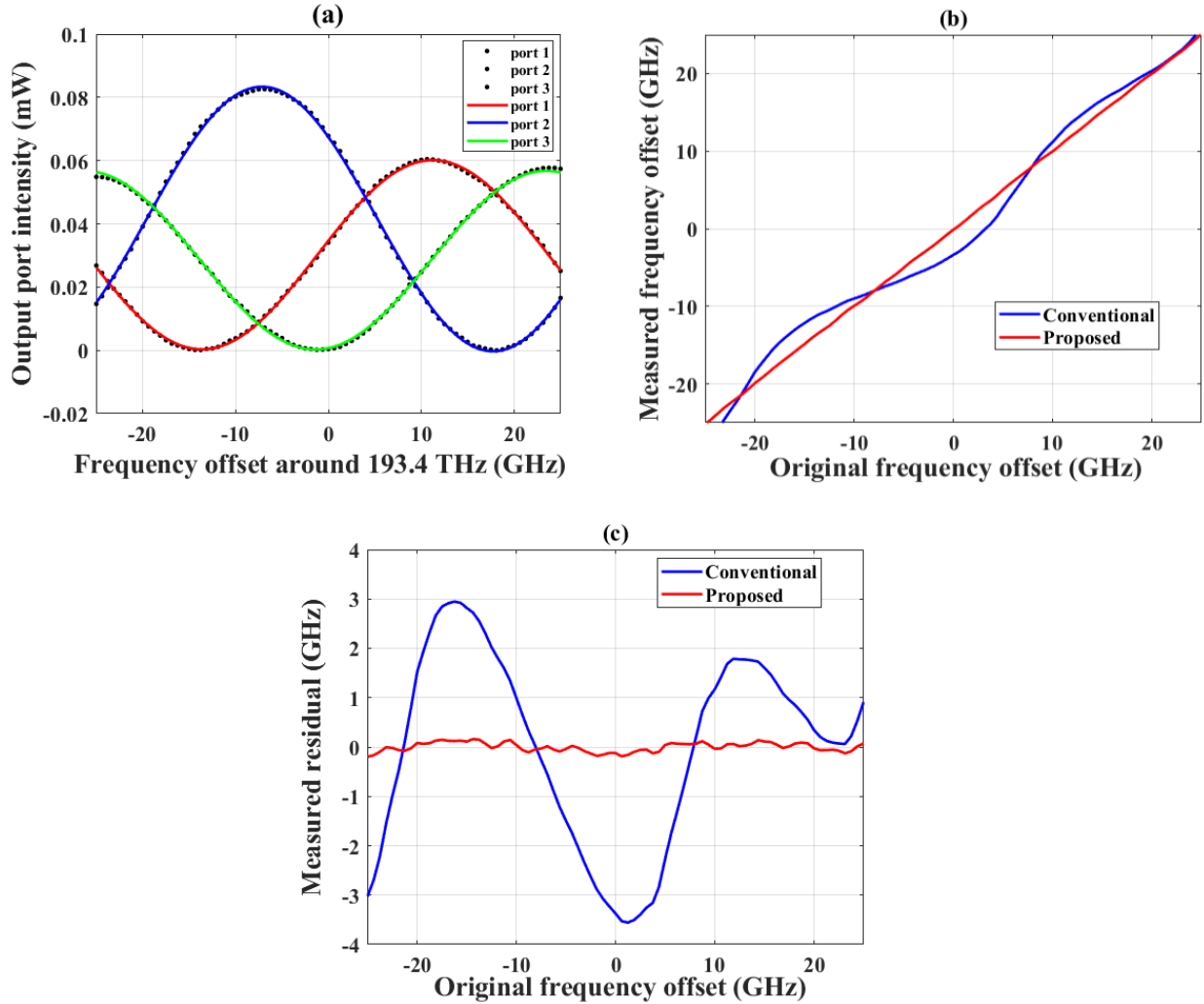

*Figure 7 (a) Recorded output port intensity (markers) from the three output ports of the 3×3 MMI coupler and the fit provided by proposed algorithm (solid); (b) frequency offset retrieved from the power sensor data by the conventional and proposed approaches versus the original frequency; (c) residual error in calculating the frequency over the desired frequency span. The reference frequency is 193.4 THz (wavelength* 1.55 *μm).*

Figure 7 depicts the experimental results associated to a frequency span of one FSR with centre vacuum wavelength 1550 nm (frequency 193.4 THz). The wavelength meter data is collected over several FSRs, but for simplicity one FSR centered at 1550 nm is selected to provide the training data set to construct the linear map ***A***. The learning algorithm is independent of the choice of training set centre wavelength or number of FSRs spanned. Once a training set is chosen, the linear mapping is optimized for the wavelength span bounded by the training set. The raw data collected from the three output ports of the $3 \times 3$ MMI coupler is shown by the markers in Figure 7(a). The result of the fit procedure is also shown by the solid line fringe pattern. An excellent fit is provided by ***A***. Figure 7(b) depicts an almost linear relationship between the original frequency recorded by the power sensor and the measured frequency due to processing via the proposed method. The conventional method is also considered which results in significant deviation of the frequency estimation. A comparison of the deviation from a linear relationship is shown in Figure 7(c), where it can be observed that the prediction of the conventional method can deviate significantly from the original frequency; the maximum residual error observed over the FSR is ~3.5 $GHz$. A improvement by over an order of magnitude is achieved by the proposed method. For the FSR considered, the residual error is limited to $\pm 0.2$ $GHz$.

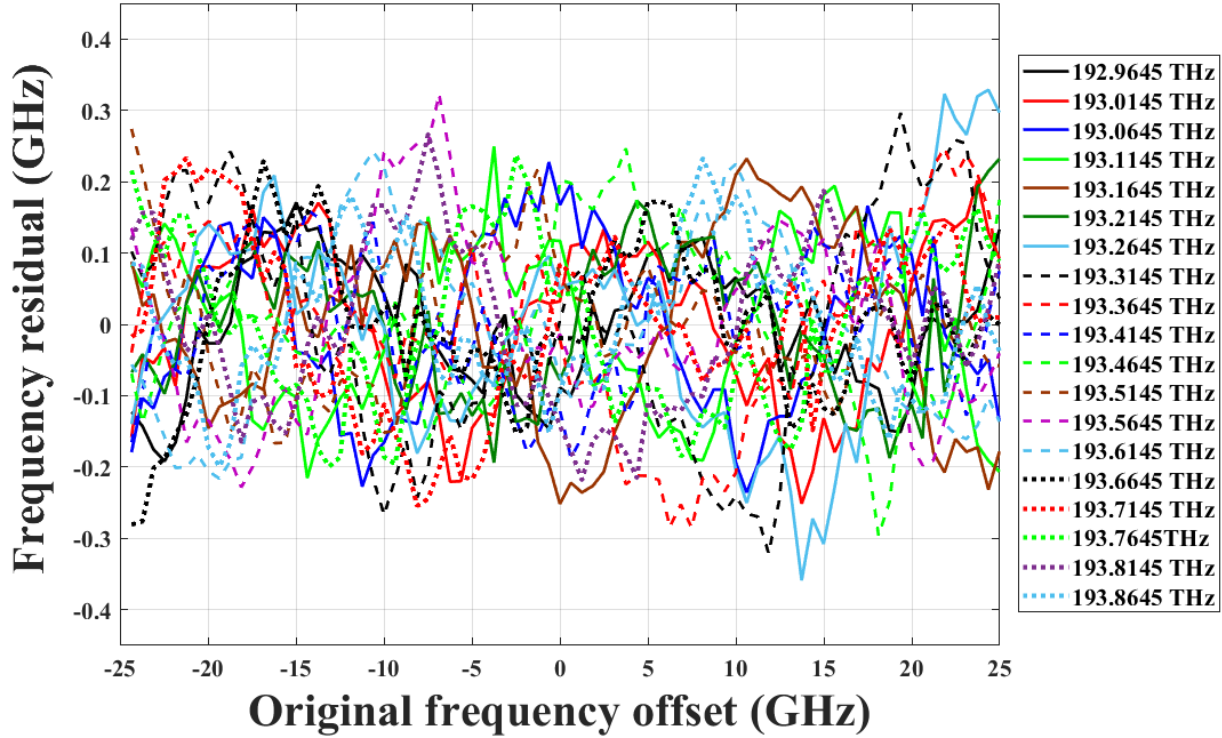

*Figure 8 Residual error in calculating the frequency over the desired frequency span for different reference frequencies*

To observe the efficacy of the method over a larger wavelength span, different training set data is taken to calibrate the system. Figure 8 shows the frequency estimation error observed for a total span of 950 GHz. Recorded data contained in one FSR around the center frequency depicted in Figure 8 are taken as the training set. After each calibration, recorded test data aligned to the respective FSR is processed by the system. It can be observed that, over the total 950 GHz span, the residual error is limited to $\pm 0.35$ $GHz$.

The experimental result shown in Figure 7 is obtained under the realistic assumption that the wavelength estimation will be performed over the same span as the training data and thus ***A*** has already been calculated. To demonstrate the generalisation ability of the learning algorithm, the linear mapping ***A*** constructed using the training data set over a specific FSR is used to retrieve the frequency using test data over an adjacent FSR. It can be observed from Figure 9 that the maximum frequency estimation of the proposed approach increases only slightly to ~0.3 GHz, which may be expected as ***A*** is not optimized for this test data set. Even with this degradation in performance, the proposed approach remains substantially superior in precision compared to the conventional method.

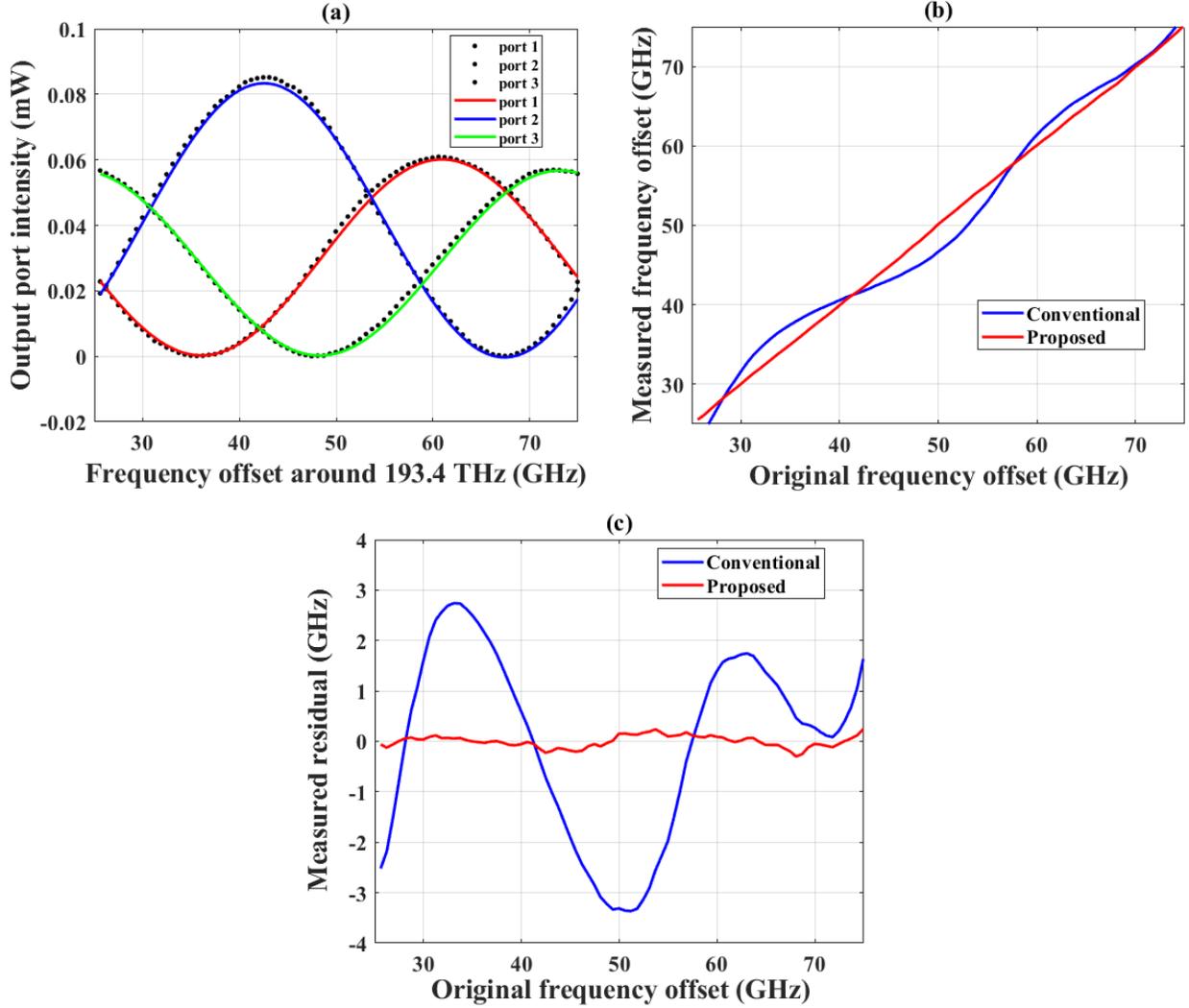

*Figure 9 (a) Recorded output port intensity (markers) from the three output ports of the 3×3 MMI coupler and the fit provided by proposed algorithm (solid); (b) frequency offset retrieved from the power sensor data by the conventional and proposed approaches versus the original frequency; (c) residual error in calculating the frequency over the desired frequency span. The reference frequency is 193.4 THz (wavelength 1.55 μm). The test data processed is extracted from the adjacent FSR to the data used for training (Figure 7)*

# V. Conclusions

This work has analysed an interferometer with three or more polyphase outputs. The theoretical analysis has informed the formulation of a machine learning and data processing method that corrects for imperfections of the interferometer components. The method has been validated by simulation and proved experimentally using a MZI based wavelength meter with fabricated on the $Si_3N_4$ photonic integration platform. The simulations demonstrate that a precision limited only by the level of random noise is attainable to the extent the model of the interferometer captures all significant impairments. The experimental observations demonstrate an order of magnitude reduction in frequency estimation error is achieved when compared to the conventional method. The maximum residual error is limited to ±0.35 GHz over a 50 GHz free-spectral range.

It has been shown that the fundamental object to be retrieved from interferometric measurements is a vector that lies on a cone in a three-dimensional object space. This vector encodes complete information on the phase imbalance and the input power of the interferometer. In the case of an interferometer that provides three or more polyphase outputs, the object space is mapped to the image (measurement) space by a linear map. In the special case of perfect components, the mapping is orthogonal. The linear map may be constructed by machine learning from a training set consisting of enough object space and image space vector pairs. In the case of a wavelength meter or frequency discriminator, the training set may be constructed from a collection of samples of the interferometer outputs paired with the frequency (vacuum wavelength) measured during a calibration phase. Formulated as a least squares minimisation problem, the learning algorithm substantially involves only linear algebra which yields an analytic solution parameterised by a single delay parameter (equivalently, free spectral range). The latter can then be found by a golden section search to minimise the residual error. The phase is retrieved by a Moore Penrose inverse and arctangent method that is invariant to source power fluctuations.

The learning algorithm corrects impairments to the couplers, delay line, photodetector sensitivity and accommodates measurement noise. The data processing algorithm is invariant to source power fluctuations and accommodates measurement noise. There is no requirement that the data processing should use the same source power as the calibration process or be normalised by a separate input power monitor or estimate. The learning algorithm is essentially linear and consequently more robust and fast compared to the nonlinear least squares method. As such the method represents a substantial advance of the prior art [16].

The principal finding of the simulations is that the proposed method will fully correct all impairments to the extent they are captured by the interferometer model (the a priori knowledge). The precision will then be limited by the noise. Since the latter is small, high accuracy is expected. The experimental results demonstrate the extrapolation of the fringes from one FSR into an adjacent FSR fit the data almost exactly even for an impaired system. The precision achieved experimentally is over one order of magnitude greater than the conventional method, but it is not as great as achieved in simulations, which indicates that performance is limited by weak impairment mechanisms not captured by the model.

The principal assumption of the model is forward transmission of a pure polarisation state with parameters that vary little over the operating frequency (i.e. over an FSR). It is concluded that one must seek phenomena involving reflections and a mixed polarisation state to explain the current limit to the precision of the wavelength meter. A learning algorithm based on a model (a priori knowledge) with numerous parameters risks overfitting the data at the expense of generalisation ability. The current wavelength meter model is parsimonious and comprehensive. Consequently, rather than increase the complexity of the model, it is preferable that interferometer meets the assumptions of the model.

Improved component and circuit designs informed by simulation and experimental investigations for polarisation and reflection control are warranted. If sufficient care is taken to avoid spurious reflections and to maintain the state of polarisation, the only deviations would be due to the slow variation over the band of the overall magnitude and phase of the fringes due to the finite bandwidth of the components and the dispersion of the waveguide. It is only errors of the cone inferred by the data processor from a data set that will propagate to subsequent phase measurements.

In that context it is advantageous to monitor calibration source power during the collection of the training data set to guard against fluctuations that misplace the object data off the circular cone. The construction of the linear map is thereby impaired which leads to an error in the phase retrieval. The resolution of this issue, if significant, is to ensure the length of the object vector is proportional to the calibration input power as measured the power monitor. In the case of $n = 3$, the $1 \times 2$ input splitter may be replaced by a $3 \times 3$ input coupler nominally identical to the $3 \times 3$ output coupler. This has the merit of a symmetrical architecture more robust to fabrication process variations and the otherwise unused central egress port of the input coupler may provide the input power monitor port. It is only necessary that the measurement is proportional to the input power; a precise value of responsivity is not required.

An important performance metric to be considered is long term stability. A rough experiment was performed using the prototype wavelength meter in which two data sets were collected with a time interval in between of several hours. One data set is used for training and the other for processing over the same FSR, and the results showed significant long-term stability. The prototype featured no input power monitoring and temperature sensor or control mechanism. So, an experimental study to assess long term stability with proper temperature control and input power monitoring is left as a future endeavor. Nevertheless, it is expected that the principal source of drift is the temperature sensitivity of the bias phase of the interferometer. This can be corrected by collecting training set data over a range of temperatures as measured by an on-chip temperature sensor. It is expected that the differences between estimated linear maps corresponding to different temperatures will be a rotation. Moreover, the rotation angle or equivalently the phase bias is expected to be linear in the temperature change as is confirmed by the experimental results reported by Chen [21]. Consequently, knowledge of the temperature coefficient is enough to compensate temperature drift.